\documentclass[aps,reprint,showpacs,superscriptaddress]{revtex4-2}

\usepackage{amssymb}
\usepackage{natbib}
\usepackage{graphicx}
\usepackage{amsmath}
\usepackage[bookmarks = false]{hyperref}
\usepackage{bm}
\usepackage[all]{hypcap}
\usepackage{colortbl}
\usepackage{booktabs}
\usepackage{braket}
\usepackage{mathrsfs}
\usepackage{multirow}
\usepackage{upgreek}
\usepackage{textcomp}
\usepackage{soul}
\usepackage{dsfont}
\usepackage[T1]{fontenc}
\usepackage{array}
\usepackage{lineno}
\usepackage{xr}
\usepackage{ulem}

\usepackage{array}
\newcommand{\PreserveBackslash}[1]{\let\temp=\\#1\let\\=\temp}
\newcolumntype{C}[1]{>{\PreserveBackslash\centering}p{#1}}
\newcolumntype{R}[1]{>{\PreserveBackslash\raggedleft}p{#1}}
\newcolumntype{L}[1]{>{\PreserveBackslash\raggedright}p{#1}}

\newcommand{\ustc}{
	\affiliation{Hefei National Research Center for Physical Sciences at the Microscale and School of Physical Sciences, University of Science and Technology of China, Hefei 230026, China}
	\affiliation{CAS Center for Excellence in Quantum Information and Quantum Physics, University of Science and Technology of China, Hefei 230026, China}
}

\newcommand{\hfnl}{
	\affiliation{Hefei National Laboratory, University of Science and Technology of China, Hefei 230088, China}
}

\newcommand{\jinan}{
	\affiliation{Jinan Institute of Quantum Technology, Jinan 250101, China}
}

\begin{document}

\title{Memory-Assisted Nonlocal Interferometer Towards Long-Baseline Telescopes}

\author{Bin Wang}\ustc\hfnl
\author{Xi-Yu Luo}\ustc\hfnl
\author{Bo-Feng Gao}\jinan
\author{Jian-Long Liu}\ustc\hfnl
\author{Chao-Yang Wang}\ustc\hfnl
\author{Zi Yan}\ustc\hfnl
\author{Qiao-Mu Ke}\ustc\hfnl
\author{Da Teng}\ustc\hfnl
\author{Ming-Yang Zheng}\jinan\hfnl
\author{Yuan Cao}\ustc\hfnl
\author{Jun Li}\ustc\hfnl
\author{Cheng-Zhi Peng}\ustc\hfnl
\author{Qiang Zhang}\ustc\hfnl\jinan
\author{Xiao-Hui Bao}\ustc\hfnl
\author{Jian-Wei Pan}\ustc\hfnl

\begin{abstract}
	\normalsize
	Quantum networks and remote quantum entanglement serve as vital future quantum communication resources with broad applicability. A key direction lies in extending the baseline of optical interferometers to enhance angular resolution in interferometric imaging. Here, by measuring a simulated thermal light field, we report the demonstration of a memory-assisted nonlocal interferometer achieving a fiber-link baseline up to 20~km while simultaneously showing its capability to compensate for a geometric delay equivalent to 1.5~km. This result demonstrates potential for enhancing the angular resolution of interferometric imaging in the optical band with delocalized single-photon entanglement, and paves the way for future application of quantum memories in astronomical observation.
\end{abstract}

\maketitle
Quantum networks~\cite{kimble_internet_2008,wehner_quantum_2018} represent a pivotal direction for the future development of quantum technologies. By employing techniques such as quantum repeater~\cite{briegel_quantumrepeater_1998}, they facilitate the interconnection and interoperability of diverse quantum devices, thereby significantly extending the scope of quantum applications. The realization of long-distance nonlocal quantum networks would enable numerous advanced use cases, including quantum cryptography~\cite{nicolas_cryptography_2002} (such as device-independent quantum key distribution (DI-QKD)~\cite{zhang_device-independent_2022,nadlinger_experimental_2022}), distributed quantum computing~\cite{jiang_distributedcomputation_2007}, and quantum sensing~\cite{gottesman_GJC_2012,komar_clock_2014}. In recent years, by incorporating quantum frequency conversion (QFC)~\cite{kumar_qfc_1990} to match signal photons to the telecom band, the fiber distance between remote matter entanglement has been extended to tens of kilometers. This progress spans diverse platforms including single atoms~\cite{van2022_33km}, cold atomic ensembles~\cite{yu2020,luo2022} and color centers~\cite{,knaut2024,stolk2024}. Notably, through the use of the Duan-Lukin-Cirac-Zoller (DLCZ) protocol~\cite{duan_long-distance_2001} and weak-field verification methods, phase coherence of Fock-state entanglement can be remotely locked and verified~\cite{liu_quantumnetwork_2024}, providing an excellent experimental platform for related applications.

One particularly promising direction lies in enhancing the angular resolution of astronomical imaging~\cite{huang_review_2025}. According to its definition, angular resolution can be improved through a longer baseline or by using shorter wavelengths. Owing to challenges such as path loss and complicated dynamical delays~\cite{hale_midinfrared_2000,monnier_2003}, nonlocal interferometry schemes in the optical regime (visible and near-infrared) that rely on direct transmission have so far been constrained to baselines not exceeding a few hundred meters~\cite{pedretti2009}, unlike their radio-frequency counterparts, which have achieved remarkable successes in recent years~\cite{akiyama_EHT_2019,akiyama_EHT_2022}. To address these limitations, D. Gottesman et al. proposed a solution in 2012, referred to as the Gottesman-Jennewein-Croke (GJC) scheme~\cite{gottesman_GJC_2012}. This approach leverages pre-established nonlocal entanglement as a phase reference to replace direct transmission, thereby effectively circumventing the issue of path loss. Furthermore, it achieves comparable Fisher information which significantly outperforms spatially local measurement strategies in weak-field measurements~\cite{tsang_FisherInfromation_2011, zhang_criteria_2025} (see Supplemental Material for details). As a result, it enables superior sensitivity and allows for observation of faint stars with enhanced accuracy. While all-optical tabletop experiment using spontaneous parametric down conversion (SPDC) source has been successfully demonstrated~\cite{brown_spdcimage_2023}, truly addressing photon loss and compensating for arbitrary geometric delays requires the incorporation of quantum memories and heralded entanglement.

In this article, we demonstrate the first nonlocal interferometer of a thermal field using two cold atomic ensembles as quantum memories. The heralded entanglement is established between the ensemble via the DLCZ protocol~\cite{duan_long-distance_2001}. A thermal light field generated through Raman scattering is then used to mimic the stellar light~\cite{mika_thermallight_2018}, which is split and separately interfered with the retrieved entanglement photons at both nodes. The interference visibility is measured via coincidence counting. We optimize parameters such as the relative intensity, quality, and indistinguishability of the thermal field in local experiments to achieve optimal visibility. Finally, by incorporating QFC and field-deployed optical fibers, we extend the baseline up to 20~km. This fiber link baseline significantly surpasses those achieved in previous experiments. The compensation capability of the memory for an equivalent 1.5~km geometric delay is also simply verified. Our experiment represents a critical step toward the construction of memory-assisted nonlocal optical interferometers. Furthermore, during the measurement of complex visibility, if the entanglement is sufficiently ideal, our experimental scheme becomes fully independent of the incident signal intensity, thereby significantly expanding its potential for practical applications in future measurements.

\begin{figure}[hbt]
	\includegraphics[width=7.2 cm]{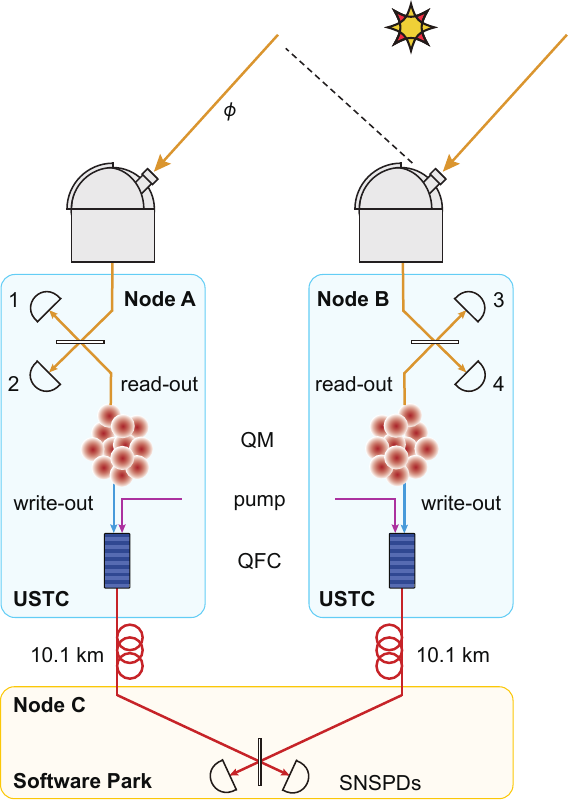}
	\caption{\label{fig:propose} The memory-assisted nonlocal interferometer. Entanglement between the memories (QM) at two nodes can be established via the DLCZ protocol. The two nodes, separated by 2~m, are housed within a laboratory at the University of Science and Technology of China (USTC). The quantum frequency conversion (QFC) and two 10.1-km deployed fibers are implemented to send the write-out photon to Node C at Hefei Software Park (located 6.5~km from Node A and B) for interference, which extends the equivalent baseline length up to 20 km. The subsequently retrieved entangled optical fields can then be separately interfered with the thermal signals locally received at Node A and B respectively. The thermal light generated through Raman scattering is used to mimic the stellar light in the experiment. By analyzing the coincidence counts between the detectors, the complex interference visibility can be determined.
	}
\end{figure}

\begin{figure}[ht]
	\includegraphics[width=8.5 cm]{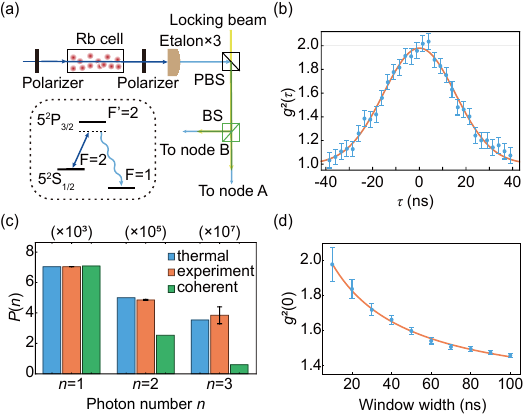}
	\caption{\label{fig:thermal} 
	Details about the thermal field generation. (a) The experiment setup for thermal field generation. (b) The measured second-order correlation function $g^{(2)}(\tau)$ with a 2.5~ns integration window width. (c) Comparison of experimental photon number distribution with theoretical models for ideal thermal and coherent light fields. (d) The measured second-order correlation function $g^{(2)}(0)$ with different integration window width. Error bars indicate one standard deviation of the photon-counting statistics.}
\end{figure} 

\begin{figure*}[hbt]
	\includegraphics[width=17 cm]{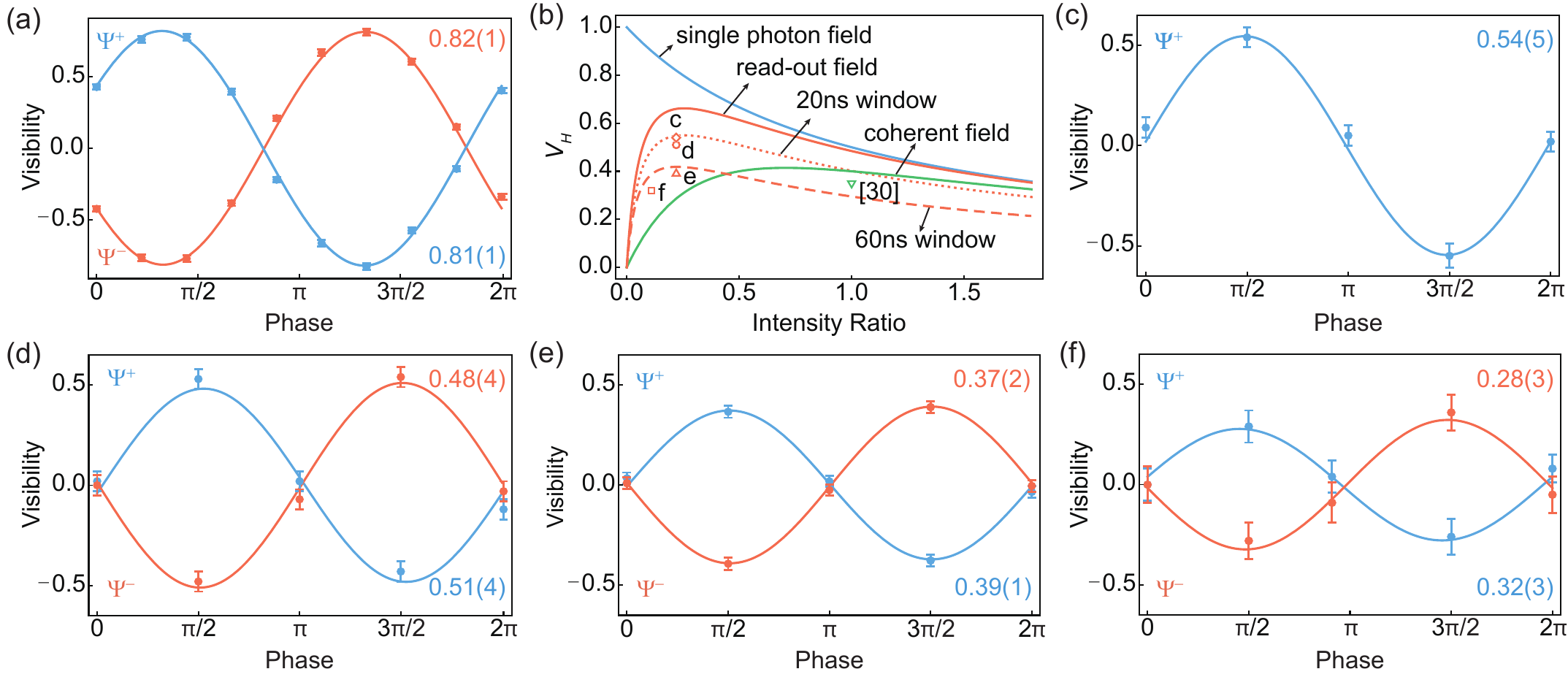}
	\caption{\label{fig:result} 
	Theory and experimental results. (a)  Verification of two-memory entanglement. (b) $V_H$ versus the intensity ratio $x$ in the main text. The solid lines correspond to the theoretical values under different auxiliary field conditions in the ideal case. The dotted (dashed) line represents the calculation results within a 20-ns (60-ns) interference window, accounting for various experimental imperfections. The orange hollow points show the experimental results presented in the subsequent figures. The green inverted triangle is the experiment result from~\cite{tang2025phase} with a visibility of 0.35. (c) Local experiment result with a 20~ns interference window. (d) The experiment result for a baseline of 20~km with a 20-ns interference window. (e) As (d) with a 60-ns interference window. (f) Interference results simulating a 5-$\upmu$s arrival time delay. All fitted visibilities are shown at the corner with the same color. Error bars indicate one standard deviation of the photon-counting statistics.}	
\end{figure*}

As a first step in demonstrating the quantum memory-assisted GJC scheme, the auxiliary entanglement between two rubidium 87 atomic ensembles at Node A and B is generated via the DLCZ protocol as shown in Fig.~\ref{fig:propose}. At each node, a weak 780 nm write pulse, blue-detuned from the atomic transition, induces a spontaneously Raman-scattered photon and (write-out) a collective atomic excitation with a small probability ($\chi\approx 0.5\%$), which form a Fock state entanglement: $\ket{\Phi_{ap}}=\ket{0_a0_p}+\sqrt{\chi}\ket{1_a1_p}$, where 0 and 1 refer to the number of atomic (subscript a) or photon (subscript p) excitations. Write-out photons from two ensembles interfere at a beam splitter (BS) to erase which-way information. A successful detection event heralds maximal entanglement between ensembles.
Leveraging ring cavity enhancement~\cite{bao2012efficient}, this entanglement can be efficiently ($\eta_{ro}\approx 26\%$) retrieved as photons, which are then used as ancillary entangled resources to participate in the interferometric measurement with the incoming stellar light. It can be described as $\rho_E$ under the Fock basis \{$\ket{00}, \ket{01}, \ket{10}, \ket{11}$\}:
\begin{equation}\label{eq:entanglement_states}
	\begin{aligned}
		\rho_E &= 
		\begin{pmatrix}
			p^E_{00} & 0 & 0 &0 \\
			0 & p^E_{01} &  \boldsymbol{d^*}  &0 \\
			0 & \boldsymbol{d} & p^E_{10} & 0 \\
			0 & 0&0 & p^E_{11}
		\end{pmatrix},
	\end{aligned}
\end{equation}
where the off diagonal element $\boldsymbol{d}=de^{i\psi}$ denotes the entanglement coherence, and can be measured by interfering these two read-out modes shown in Fig.~\ref{fig:result}a, yielding concurrence $\mathcal{C} (\Psi^+)= 0.127\pm0.004$ and $\mathcal{C}(\Psi^-) = 0.129\pm0.004$~\cite{Chou_concurrence_2005}. 

To emulate genuine thermal stellar light in the laboratory, previous studies predominantly utilized rotating ground-glass diffusers to generate pseudo-thermal light fields~\cite{brown_spdcimage_2023,tang2025phase}, which still contain residual periodicity from the glass rotation~\cite{lee2020}. Here we implement an alternative approach~\cite{mika_thermallight_2018} to eliminate this defect as shown in Fig.~\ref{fig:thermal}a. A rubidium vapor is excited by a pre-polarized 780~nm laser pulse, which induces an omnidirectional spontaneous photon emission. A small fraction of Doppler-shift-free scattered photons, which can serve as the required thermal light field, propagate collinearly with the excitation beam~\cite{dussaux2016}. A filtering module is constructed to suppress various noise components in this field, including residual laser light and Rayleigh scattering components. This module consists of a polarizer with an extinction ratio up to 50~dB for primary laser rejection, and the cascaded triple etalons with an extinction ratio of about 30~dB per stage to remove the remaining scattered photons with a 6.8~GHz frequency difference. The module has an overall efficiency of 30\% and an extinction ratio of 130~dB. Then we measure the statistical properties of the output field and the fitted second-order correlation function $g^{(2)}(0)=1.98\pm0.01$ as shown in Fig.~\ref{fig:thermal}b, with the photon number statistics shown in the Fig.~\ref{fig:thermal}c. These results collectively demonstrate that the field serves as a good approximation to a thermal light source and is suitable for use in our interferometric experiments.

Furthermore, as shown in Fig.~\ref{fig:thermal}b, the measured second-order correlation function $g^{(2)}(\tau)$ with a 2.5~ns window width reveals the coherence time of $15.4\pm0.3$~ns~\cite{dussaux2016}. Given the short coherence time, the integration window width must be minimized. Otherwise, as illustrated in Fig.~\ref{fig:thermal}d, an excessively wide detection window yields reduced $g^{(2)}(0)$, which directly degrades the final interference visibility.  

As shown in the Fig.~\ref{fig:thermal}a, the output light field is combined with the phase-locking beam by a polarizing beam splitter (PBS) and split by a BS to form the path entanglement state described as $\rho_S$: 
\begin{equation}\label{eq:thermal_states}
	\begin{aligned}
		\rho_S &=
		\begin{pmatrix}
			p^S_{00} & 0 & 0 &0 \\
			0 & p^S_{01} & \boldsymbol{g^*}  &0 \\
			0 & \boldsymbol{g} & p^S_{10} & 0 \\
			0 & 0&0 & p^S_{11}
		\end{pmatrix},
	\end{aligned}
\end{equation}
where the off diagonal term $\boldsymbol{g}=ge^{i\phi}$ of $\rho_S$ is known as the complex visibility, a function of the baseline vector between the telescopes that needs to be measured by the interferometers. According to the original GJC protocol, we implement three-fold coincidence counting (a click at each node) as valid events, and the complex visibility $\boldsymbol{g}$ can be extracted by coincidence visibility if we scan the phase $\psi$ of the auxiliary entanglement (see Supplemental Material for detail derivation):
\begin{equation}\label{eq:visibility}
	\begin{aligned}
		V(\psi,\phi)&=\frac{N_{13}+N_{24}-N_{14}-N_{23}}{N_{13}+N_{24}+N_{14}+N_{23}}\\
		&=\frac{2\cdot P_E(1)\cdot P_S(1)\cdot d\cdot g\cdot \cos(\psi-\phi)\cdot \xi}{P_E(0)P_S(2)+P_E(2)P_S(0)+2P_E(1)P_S(1)},	
	\end{aligned}
\end{equation}
where $N_{ij}$ stands for the coincidence clicks between the $i$ detector in Node A and the $j$ detector in Node B as shown in Fig.~\ref{fig:propose}, and $P_{S(E)} (n)$ represents the probability of simultaneously detecting $n$ photons from stellar (auxiliary entangled) light field. The mode mismatch effect is quantified by $\xi$, which can be estimated and optimized through Hong-Ou-Mandel (HOM) interference measurements. A cornerstone of the GJC scheme lies in its utilization of auxiliary entangled optical fields to provide a stable and tunable phase reference. Phase-locking techniques become indispensable when interferometer baseline lengths extend to tens or even hundreds of kilometers. In this experiment, we dynamically locked three interferometers to control the phase of single-photon entangled states. Detailed optimizations of interference and phase-locking schemes are provided in the Supplemental Material.

Various theoretical factors affecting the visibility can be identified in Eq.~\ref{eq:visibility}, where the primary limiting one is multi-photon events. Since neither light field constitutes a perfect single-photon source, detector responses triggered by multi-photon events, such as two thermal photons or two auxiliary light field photons, are erroneously registered as valid coincidence counts, thereby reducing the visibility. Isolating higher-order contributions' effect on visibility, $V_{H}$ is derivable from Eq.~\ref{eq:visibility}.
\begin{equation}\label{eq:highordervis}
	\begin{aligned}
		V_{H}&\approx
		\frac{2P_E(1)P_S(1)}{P_E(0)P_S(2)+P_E(2)P_S(0)+2 P_E(1)P_S(1)}\\
		&=\frac{2x}{x^2\cdot g^{(2)}_S(0)+g^{(2)}_E(0)+2x},
	\end{aligned}
\end{equation}
$x=P_S(1)/P_E(1)$ is the brightness ratio between the thermal field and auxiliary light field. There exists an optimal value of brightness ratio $x$ that maximizes the interference visibility as the orange solid line shown in Fig.~\ref{fig:result}b with $g^{(2)}_S = 1.98\pm0.01$ for the thermal field and $g^{(2)}_E = 0.13\pm0.08 $ for the read-out field. Additionally, the figure includes comparative theoretical curves for both ideal single-photon entanglement and coherent light fields. It can be observed from the plot that, due to the presence of multi-photon events in the generated entanglement, the theoretical visibility of our experiment is lower than that achievable with ideal single-photon entanglement. Nevertheless, it still surpasses the performance possible with coherent light fields. Furthermore, our approach achieves optimal visibility with a lower-intensity thermal probe field, highlighting the advantage of the entanglement-based scheme.

We initially perform the experiments with the write-out path maintained locally. The source brightness is calibrated to be $P_S(1)=0.06$ according to the optimal ratio $x=0.22$ and a retrieve efficiency $\eta_{ro}\approx 26\%$, following Eq.~\ref{eq:highordervis}. Under identical 20~ns window conditions, this configuration achieves visibility of $0.54\pm0.05$ as shown in Fig.~\ref{fig:result}c. This experimental result agrees well with the theory (dotted line in the Fig.~\ref{fig:result}b) that includes other experimental imperfections. Only one curve is shown because the other port of the write-out BS is dedicated to phase locking in the local setup.

As shown in the Fig.~\ref{fig:propose}, the path length of the write-out photons principally determines the baseline of the entanglement-assisted interferometer. To extend the baseline of the interferometer to 20~km, we leverage QFC technology to convert the write-out photon from 780~nm to 1522~nm~\cite{luo2025entanglingquantummemories420}. The write-out photons are then guided to interfere at Node C through a field-deployed fiber of 10.1~km. The visibility under this configuration is $0.51\pm0.04$ as shown in Fig.~\ref{fig:result}d, which shows no significant deviation from the local result above. Meanwhile, to assess the coherence degradation effect, we analyzed the same dataset using a 60~ns window that encompasses the full wavepacket but far exceeds the coherence time. The fitted visibility is $0.39\pm0.01$ as shown in Fig.~\ref{fig:result}e.

In previous experiments, the thermal field signals always arrived at the interferometer simultaneously. However, in practical applications, non-zero differences in wavefront arrival times are often inherent. To functionally verify that our memory system can be applied to such scenarios by retrieving the entanglement at different time, we add a 1~km fiber spool to one arm of the split thermal light field to simulate this delay. A BS is placed in the other arm to balance the intensity. Simultaneously, the corresponding memory is read out with 5~$\upmu$s relative delay to interfere with it. The measured visibility of $0.32\pm0.03$ with 60~ns window, though lower than the delay-free case, confirms the memory's capability for this more realistic situation. The observed reduction primarily stems from suboptimal brightness ratios $x =0.11$ caused by transmission loss in the 1~km fiber spool and BS. Further brightness enhancement is constrained by excitation laser saturation and losses introduced during wavepacket shaping. A comprehensive analysis of all conceivable experimental imperfections is provided in the Supplemental Material.

In this work, we employed entanglement between quantum memories separated by 20~km of optical fiber to construct a nonlocal optical interferometer. The complex visibility of the thermal field collected from a rubidium vapor cell was measured, yielding the highest interference visibility with the longest baseline reported to date in entanglement-based experiments. Notably, even with imperfections such as multiple excitations present in the prepared ancillary entanglement, the sensitivity performance we achieved still surpasses the theoretical limit attainable by the coherent field sharing method~\cite{tsang_FisherInfromation_2011,zhang_criteria_2025,tang2025phase}, enabling higher precision for an equal number of measurements. More fundamentally, the Fisher information, which directly reflects detection sensitivity, indicates that the entanglement-based nonlocal interferometry can achieve significantly higher information extraction in a single detection, leading to substantially enhanced estimation precision. Although this experiment employs delay-choice during entanglement generation, the use of spin-wave freezing techniques~\cite{jiang_spinfrezzing_2016,luo2022,liu_quantumnetwork_2024} makes it feasible to extend memory lifetime beyond the round-trip signal transmission time required by the current baseline length, thereby enabling operation in an event-ready regime. Furthermore, with future integration of long-lived quantum memories possessing sub-second lifetime~\cite{wang_longlife_2021}, the interferometer's baseline can be straightforwardly extended to hundreds of kilometers.

If future implementations achieve true spatial separation, our memory-assisted interferometer has the potential to reach an angular resolution of $\lambda/B \approx 8~\upmu$as, rivaling the capability of the Event Horizon Telescope (EHT)~\cite{akiyama_EHT_2019}. Achieving full separation between nodes is no longer a major difficulty, several experiments~\cite{van2022_33km,knaut2024,stolk2024} have already yielded promising results in this regard. Particularly, our experiment can be implemented through the integration of remote phase-locking techniques~\cite{liu_quantumnetwork_2024}. Meanwhile, the implementation of multiplexing techniques~\cite{collins2007,simon_multimode_2007,pu2017,heller_multiplexing_2020,wang_multiplexing_2023,zhang_multiplexing_2024} and binary-encoded schemes~\cite{khabiboulline2019,Khabiboulline2019_array,Czupryniak_qubitcircuits_2023} can effectively increase the entanglement generation rate and reduce the required auxiliary entanglement consumption respectively, which makes the experiment more viable for practical deployment. The integration and further development of high-efficiency, low-noise, multi-band QFC technologies will enable the simultaneous and efficient utilization of different spectral components present in stellar light~\cite{Khabiboulline2019_array}, thereby enhancing overall resource utilization efficiency. Additionally, in the future, The deployment of quantum networks incorporating technologies such as quantum repeaters~\cite{briegel_quantumrepeater_1998}, quantum error correction~\cite{huang_errorcorrect_2022} and distillation~\cite{wang_Distillation_2025} holds the potential to significantly extend achievable baseline lengths, thereby  enabling the establishment of larger-scale optical interferometer arrays~\cite{gottesman_GJC_2012}.

This research was supported by the Quantum Science and Technology-National Science and Technology Major Project (No.~2021ZD0301104, No.~2021ZD0300802, No.~2024ZD0300301), National Key R\&D Program of China (No.~2020YFA0309804), Anhui Initiative in Quantum Information Technologies, National Natural Science Foundation of China (No.~T2525008, No.~T2125010, No.~92476203), Chinese Academy of Sciences. 

\textit{Note added.}---During the preparation of this manuscript, we became aware of a related experiment~\cite{stas2025_interferometry} that realizes entanglement assisted nonlocal optical interferometry with color centers, achieving a fiber baseline up to 1.55 km.

\setcounter{figure}{0}
\setcounter{table}{0}
\setcounter{equation}{0}

\onecolumngrid

\global\long\def\theequation{S\arabic{equation}}
\global\long\def\thefigure{S\arabic{figure}}
\global\long\def\thetable{S\arabic{table}}
\renewcommand{\arraystretch}{0.6}

\newpage

\newcommand{\msection}[1]{\vspace{\baselineskip}{\centering \textbf{#1}\\}\vspace{0.5\baselineskip}}

\msection{SUPPLEMENTAL MATERIAL}

\section{Fisher information about geometry phase}
The main function of the Very Long Baseline Interferometry (VLBI) is to measure the phase and the visibility of the complex visibility of the stellar light. Considering the case average photon number $\epsilon \ll 1$, as is common for interferometry with high optical frequency, the density operator can be approximated in the photon-number basis:
\begin{equation}
	\begin{aligned}
		\rho_S =(1-\epsilon)\ket{0,0} \bra{0,0}+\frac{\epsilon}{2}[\ket{0,1}\bra{0,1}+\ket{1,0}\bra{1,0}+\boldsymbol{g}\ket{1,0}\bra{0,1}+\boldsymbol{g}^*\ket{0,1}\bra{1,0}]+\mathcal{O}(\epsilon^2).
	\end{aligned}
\end{equation}
According to~\cite{tsang2011}, the Fisher information of the complex visibility $\boldsymbol{g}=ge^{i\phi}$ for direct detection can be calculated after an adjustable phase shift $\delta$ added:
\begin{equation}
 F=\frac{\epsilon}{1-\mathrm{Re}(ge^{i(\phi-\delta)})^2}\begin{pmatrix}
			\cos^2(\phi-\delta) & \sin(\phi-\delta)\cos(\phi-\delta) \\
			\sin(\phi-\delta)\cos(\phi-\delta) & \sin^2(\phi-\delta) \\
		\end{pmatrix}.
\end{equation}
The eigen values of $F$ is $\lambda_1=0$ and $\lambda_2=\frac{\epsilon}{1-\mathrm{Re}(ge^{i(\phi-\delta)})^2}$, so the trace norm of Fisher infromation is $\Vert F\Vert=\lambda_1+\lambda_2 \geq \epsilon$, which scales linearly with the average photon number (the Fisher information for the GJC scheme has in theory the same expression but reduced by a factor of 2). Compared to the local measurement schemes bounded by a Fisher information ($\Vert F\Vert\leqslant  \epsilon^2$), both direct interferometry and the GJC approach offer a substantial advantage in low-brightness source detection ($\epsilon\ll1$). In other words, spatially local measurements are fundamentally much less efficient than our  methods in extracting coherence information from weak-thermal-light interferometry.

Considering the imperfections in practical measurements, the experimentally obtained Fisher information can be expressed as:
\begin{equation}
 F=\frac{\eta\epsilon}{2(1-\mathrm{Re}(Ve^{i(\phi-\delta)})^2)}\begin{pmatrix}
			\cos^2(\phi-\delta) & \sin(\phi-\delta)\cos(\phi-\delta) \\
			\sin(\phi-\delta)\cos(\phi-\delta) & \sin^2(\phi-\delta) \\
		\end{pmatrix},
\end{equation}
where $\eta$ is the retrieval efficiency of the entanglement, and $V$ is the visibility amplitude measured in the experiment. It is easy to get the upper bound: $\Vert F\Vert\leqslant\frac{\eta\epsilon}{2(1-V^2)}$. A higher visibility amplitude ($V$) leads to a more sensitive measurement, which is the key motivation for this work to optimize the experimental setup.

\section{Measurement Protocol}

\begin{figure}[htb]\label{fig:measurement}
	\includegraphics[width=0.65\textwidth]{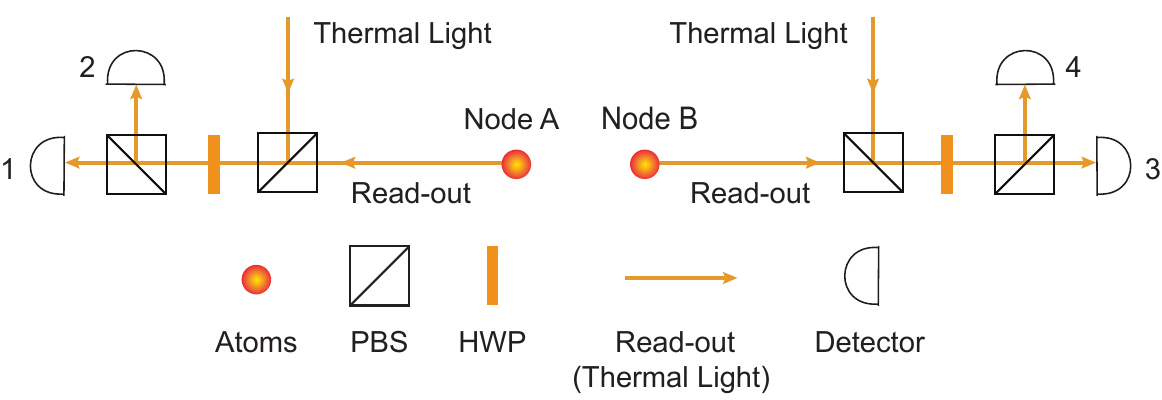}
\caption{Measurement protocol.
}
\end{figure}

Here we briefly introduce the protocol we use to measure the complex visibility of thermal light and the derivation process  of the Eq.~\ref{eq:visibility} in the main text. As shown in the Fig.~\ref{fig:measurement}, our actual experimental setup differs slightly from the Fig.~\ref{fig:propose} in the main text. In practice, we combined the thermal light field and the read-out field using a PBS, and then employed a HWP placed at $45^{\circ}$ along with another PBS to achieve interference and detection. The final results are identical to those obtained by directly using a BS. The decision not to use a BS was primarily based on the convenience of phase-locking, as well as the difficulty in ensuring optimal balance in a BS. With the density matrices in the main text, the state after mixing reads
\begin{equation}
	\rho=\rho_E\otimes\rho_S,
\end{equation}
the HWP placed at $45^{\circ}$ and the PBS forms a detection module. We register the events where only one detector clicks at each node. For instance, the projection operator $\hat{W}_{13}$ corresponding to the joint response of the detector 1 and 3 in Fig.~\ref{fig:measurement} can be expressed as:
\begin{equation}
	\hat{W}_{13}=(I-\ket{0}\bra{0})_1\otimes\ket{0}\bra{0}_2\otimes(I-\ket{0}\bra{0})_3\otimes\ket{0}\bra{0}_4,
\end{equation}
the probability of the measured response is given by the expectation value of this projection operator, yielding
\begin{equation}
	\begin{aligned}
	N_{13}&=Tr[\rho\hat{W}_{13}]\\
	&=\frac{1}{4}\{P_E(2)\cdot P_S(0)+P_E(0)\cdot P_S(2)+2P_E(1)\cdot P_S(1)+2P_E(1)\cdot P_S(1)\cdot d\cdot g \cdot \cos(\psi-\phi)\},
	\end{aligned}
\end{equation}
where $P_{S(E)}(n)$ represents the probability of simultaneously detecting $n$ photons from stellar (auxiliary entangled) light field.
Following the same procedure for the other three projection operators, we arrive at the visibility formula Eq.~\ref{eq:visibility} provided in the main text:
\begin{equation}
\begin{aligned}
	V(\psi,\phi)&=\frac{Tr[\rho(\hat{W}_{13}+\hat{W}_{24}-\hat{W}_{14}-\hat{W}_{23})]}{Tr[\rho(\hat{W}_{13}+\hat{W}_{24}+\hat{W}_{14}+\hat{W}_{23})]}\\
	&=\frac{N_{13}+N_{24}-N_{14}-N_{23}}{N_{13}+N_{24}+N_{14}+N_{23}}\\
	&=\frac{2\cdot P_E(1)\cdot P_S(1)\cdot d\cdot g\cdot \cos(\psi-\phi)}{P_E(0)P_S(2)+P_E(2)P_S(0)+2P_E(1)P_S(1)}.
\end{aligned}
\end{equation}

\section{Details of phase stabilization}
To ensure stable interference, this experiment required active phase locking of three interferometers. The details of the phase-locking systems are described individually in the following subsections.
\begin{figure}[htb]\label{fig:phase_locking}
	\includegraphics[width=0.9\textwidth]{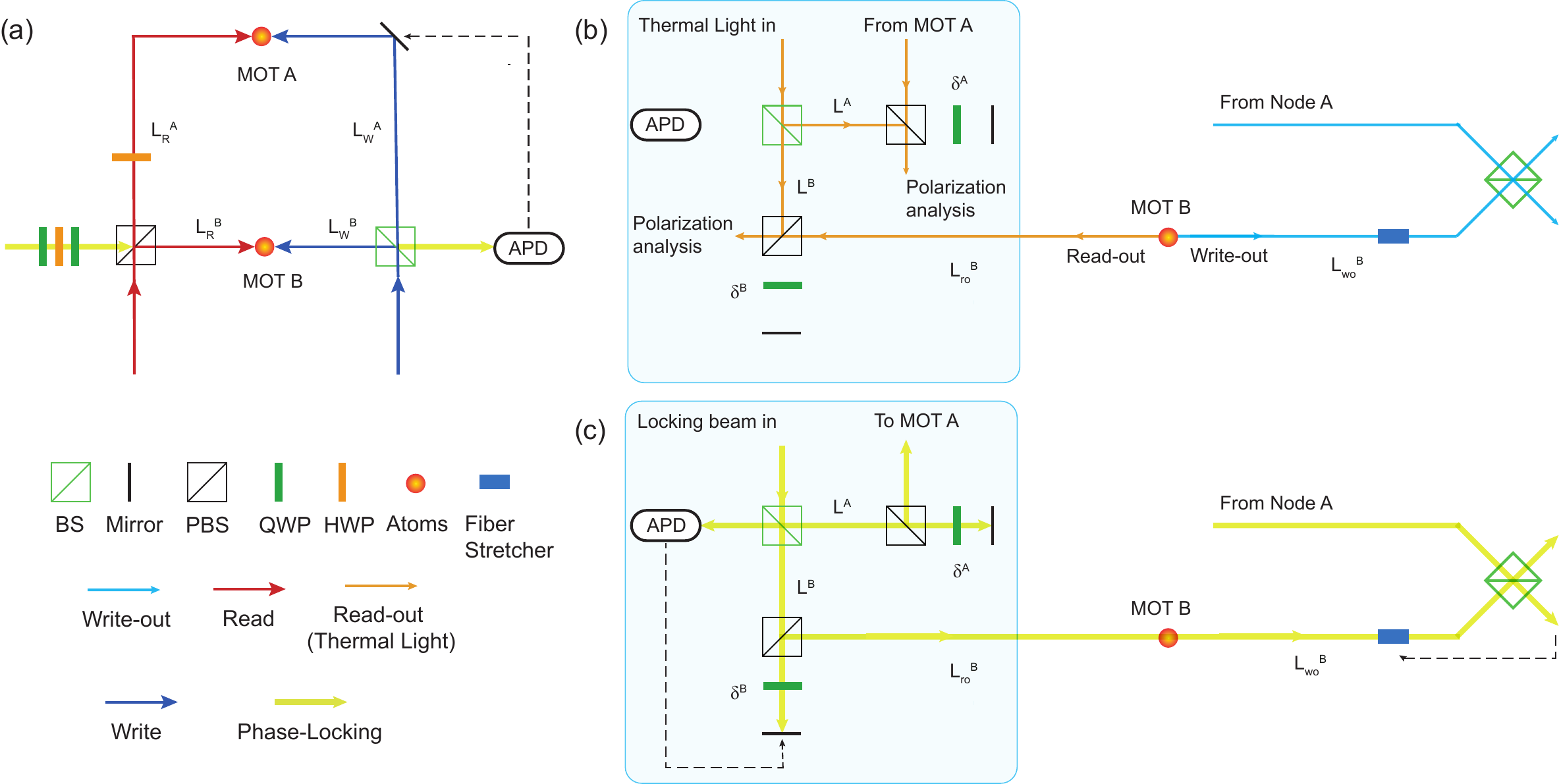}
\caption{Phase locking scheme details. (a) The local write-read interferometer (named WRI) covers paths of write and read paths. The relative locking phase between two arms can be adjusted by a sandwich configuration of wave plates. It's also the way we introduce a controllable phase shift in auxiliary entanglement verification and GJC scheme demonstration. (b) Interferometer configuration covering the write-out, read-out, and thermal light paths. (c) The two locking-beam interference loops used to stabilize the optical paths in (b). The first loop (named THI, schematically indicated within blue box) is similar in configuration to a Michelson interferometer. The locking beam is reflected by the mirror at the bottom, passing twice through the transmission path of the thermal field and a short extra optical path. This arrangement is primarily used to stabilize the relative phase of the thermal field transmission path. A portion of the reflected beam from the mirror is reflected via a PBS (with relative intensity adjusted by the QWP) to cover both the read-out and write-out paths, and subsequently interferes at the detection node for feedback, thereby stabilizing the phase of the overall read-out-write-out path (named WOI). Similar to Ref.~\cite{luo2025}, we also implement dual-wavelength locking for write-out path, which includes the field-deployed fiber.
}
\end{figure}

\subsection{Phase locking of the thermal entanglement}

The interferometer THI in Fig.~\ref{fig:phase_locking}c is similar as a Michelson interferometer. A piezo-electric transducer control the mirror at the bottom to actively stabilize the phase between two arms of THI:
\begin{equation}\label{eq:first_interferometer}
2k_{\rm p}(L^{B}-L^{A}+\delta^{B}-\delta^{A}) = \phi_\mathrm{th},
\end{equation}
where $k_{\rm p}$ is the wave number of the locking beam. Meanwhile the weak vertical polarized thermal-ﬁeld entangled state can be expressed as:
\begin{equation}\label{eq:t_entanglement}
\ket{\Phi}_\mathrm{th}=\ket{V0}_{\rm AB}+\exp[ik_\mathrm{th}(L^{B}-L^{A})]\ket{0V}_{\rm AB}.
\end{equation}

\subsection{Phase locking of auxiliary entanglement}
We review the process of the auxiliary single photon entanglement creation. A write pulse induces a collective excitation and a write-out scattered photon which form a Fock state entanglement between two memories. Then the atomic state is retrieved as a read-out photon for interfering. In the following derivation, we omit the frequency conversion process, which does not introduce additional complications since the two pump beams originate from the same laser.
\subsubsection{Write process}

Each node emits write pluses to excite its local atomic ensemble, which subsequently radiates write-out photons directed to the central node for interference. Upon successful heralding, the ensembles are projected into the entangled state:
\begin{equation}
	\ket{\Phi}_{\rm atom}=\exp[iW_{\rm A}]\ket{SG}_{\rm AB}\pm\exp[iW_{\rm B}]\ket{GS}_{\rm AB},
\end{equation}
here $\ket{S(G)}$ stands for the excitation (ground) sate of ensembles and
$W_{\rm A}=\Phi_{\rm w}^{A}-(k_{\rm w}L^{A}_{\rm w}+k_{\rm wo}L^{A}_{\rm wo})+\phi_{\rm woro}$,
$W_{\rm B}=\Phi_{\rm w}^{B}-(k_{\rm w}L^{B}_{\rm w}+k_{\rm wo}L^{B}_{\rm wo})$, where $\phi_{\rm woro}$ represents a active compensation phase applied to the fiber stretcher in one arm of WOI in Fig.~\ref{fig:phase_locking}c, satisfying the condition:  
\begin{equation}\label{eq:woro}
	k_{\rm p}[L^{B}+2\delta^{B}+L^{B}_{\rm ro}+L^{B}_{\rm wo}-(L^{A}+2\delta^{A}+L^{A}_{\rm ro}+L^{A}_{\rm wo})]=\phi_{\rm woro},
\end{equation} where $k_\mathrm{p}\approx k_{\rm wo}\approx k_{\rm ro}=k_{\rm th}$.

\subsubsection{Read process}
The atomic state is retrieved by read pluses. The horizontal polarized read-out single photon entangled state at PBS can be expressed as:
\begin{equation}\label{eq:a_entanglement}
	\ket{\Phi}_{\rm ro}=\exp[i(R_{\rm A}+W_{\rm A})]\ket{H0}_{\rm AB}\pm\exp[i(R_{\rm B}+W_{\rm B})]\ket{0H}_{\rm AB},
\end{equation}
here $R_{\rm A(B)}=\Phi_{\rm r}^{A(B)}-k_{\rm r}L_{\rm r}^{A(B)}-k_{\rm ro}L_{\rm ro}^{A(B)}$. In the experimental setup, since both atomic ensembles share the same write and read laser sources, the initial phase of lasers, $\Phi_{\rm w}^{A(B)}, \Phi_{\rm r}^{A(B)} $, can be neglected. The optical paths for writing and reading collectively form an equal-arm interferometer (named WRI) in Fig.~\ref{fig:phase_locking}a. By adjusting the half wave plates (HWP) in the sandwich configuration in Fig.~\ref{fig:phase_locking}a, we can control the locking point of this interferometer, yielding:
\begin{equation}\label{eq:wr}
	k_{\rm w}L^{B}_{\rm w}+k_{\rm r}L^{B}_{\rm r}-k_{\rm w}L^{A}_{\rm w}-k_{\rm r}L^{A}_{\rm r}\approx k_{\rm wr}(L^{B}_{\rm w}+L^{B}_{\rm r}-L^{A}_{\rm w}-L^{A}_{\rm r})=\phi_{\rm wr},
\end{equation}
where $k_{\rm wr}(\approx k_{\rm w}, k_{\rm r})$ is the wave number of the locking beam. We substitute Eq.~\ref{eq:wr} and Eq.~\ref{eq:woro} into Eq.~\ref{eq:a_entanglement}, it gives
\begin{equation}\label{eq:as_entanglement}
	\ket{\Phi}_{\rm ro}=\ket{H0}_{\rm AB}\pm\exp[ik_{\rm p}(2\delta^{B}-2\delta^{A}+L^{B}-L^{A})-(\phi_{\rm woro}+\phi_{\rm wr})]\ket{0H}_{\rm AB}.
\end{equation}

\subsection{Total phase of the two optical fields}
After the interference between the thermal field and the read-out field, we select the case where one photon is detected at each node, $\ket{\Phi}_{\rm th}\ket{\Phi}_{\rm ro}$ becomes:
\begin{equation}
	\ket{\Phi}_{\rm POL}=\ket{HV}\pm\exp[i(\phi_{\rm th}-\phi_{\rm woro}-\phi_{\rm wr})]\ket{VH}.
\end{equation}
The three phases here can be controlled and stabilized via the three interferometers described above, thereby granting the ultimately measured interference visibility a stable phase reference. Tab.~\ref{tab:symbols} summarizes the definitions of all mathematical symbols used in the derivations.

\begin{table*}[h]
	\centering
	\caption{A summary of notations and their descriptions in the derivation. We use the superscript A, B for referring to the channels from node A and node B respectively.}
	\begin{tabular}[t]{C{2cm} C{10cm}}
		\toprule
		Notations & Description \\
		\midrule
		$k_{\rm th}$& The wave number of thermal light.\\
		$k_{\rm p}(k_{\rm wr})$& The wave number of phase locking light.\\
		$k_{\rm w}$& The wave number of write light.\\
		$k_{\rm wo}$& The wave number of write out light.\\
		$k_{\rm r}$& The wave number of read light.\\
		$k_{\rm ro}$& The wave number of read out light.\\
		$L^{A/B}$& The A(B) path length of thermal field.\\
		$\delta^{A/B}$& The A(B) excess path length of the phase locking beam relative to the signal beam.  \\
		$\Phi_{\rm w}^{A/B}$& The A(B) initial phase of the write light source.\\
		$\Phi_{\rm r}^{A/B}$& The A(B) initial phase of the read light source.\\
		$L_{\rm w}^{A/B}$& The A(B) path length of write beam.\\
		$L_{\rm wo}^{A/B}$& The A(B) path length of write out beam.\\
		$L_{\rm r}^{A/B}$& The A(B) path length of read beam.\\
		$L_{\rm ro}^{A/B}$& The A(B) local path length of read out beam.\\
		\bottomrule
	\end{tabular}
	\label{tab:symbols}
\end{table*}

\section{Experimental imperfection of Visibility}
As derived in the main text, the interference visibility is governed by three primary factors: the multi-photon term, the coherence degree of the two optical fields, and the phase instability after phase locking. Furthermore, imperfect mode matching between the fields introduces additional visibility reduction. We now analyze each factor systematically.

\subsection{Decrease of SNR (Signal to Noise Ratio)}
The experiment record three-fold coincidence clicks: one click at the write-out detector and one click at each node. Since it is impossible to determine whether the write-out detector response originated from a genuine write-out photon or noise, the presence of the latter degrades the interference visibility:
\begin{equation}
	\begin{aligned}
		V&=\frac{(N_{\rm max}+n/2)-(N_{\rm min}+n/2)}{(N_{\rm max}+n/2)+(N_{\rm min}+n/2)}\\
		&=\frac{N_{\rm max}-N_{\rm min}}{N_{\rm max}+N_{\rm min}+n}\\
		&=V_0 \cdot \frac{co_{\rm wo}}{co_{\rm wo}+co_{\rm noise}}\\
		&=V_0\cdot V_{\rm SNR},
	\end{aligned}
\end{equation}
$N$ denotes signal counts and $n$ represents noise counts, $co_{\rm wo}\approx p_{\rm wo}*\eta_{\rm ro}$ ($co_{\rm noise}\approx p_{\rm noise}*p_{\rm ro}$) means coincidence probability of the real write out - read out (noise) event. The decrease of $V_{\rm SNR}$ can be estimated by $\mathrm{SNR}=\frac{p_{\rm wo}}{p_{\rm noise}}$:
\begin{equation}
	\begin{aligned}
		V_{\rm SNR}&=\frac{co_{\rm wo}}{co_{\rm wo}+co_{\rm noise}}\\
		&=\frac{p_{\rm wo}*\eta_{\rm ro}}{p_{\rm wo}*\eta_{\rm ro}+p_{\rm noise}p_{\rm ro}}\\
		&=\frac{\mathrm{SNR}*\eta_{\rm ro}}{\mathrm{SNR}*\eta_{\rm ro}+p_{\rm ro}}.
	\end{aligned}
\end{equation}
After applying the parameters in our experiments, the decreases that $V$ suffers from SNR are listed in Tab.\ref{tab:SNR}
\begin{table}[h]\label{tab:SNR}
	\centering
	\caption{A summary of SNR and $V_{\rm SNR}$.}
	\begin{tabular}{C{3cm} C{2cm} C{2cm}}
		\toprule
		& SNR & $V_{\rm SNR}$ \\
		\midrule
		20~km experiment & $3.23\pm 0.07$ & $0.994\pm 0.030$ \\
		\midrule
		local experiment & $8.80\pm 0.19$ & $0.998\pm 0.030$ \\
		\bottomrule
	\end{tabular}
\end{table}

\subsection{Higher order events}
The maximum visibility $V_H$ in the main text is the ideal maximum visibility defined by isolating the multi-photon contribution while neglecting other noise sources:
\begin{equation}\label{eq:idealvis}
	\begin{aligned}
		V_{H}&\approx
		\frac{2P_{E}(1)P_{S}(1)}{P_{E}(0)P_{S}(2)+P_{E}(2)P_{S}(0)+2P_{E}(1)P_{S}(1)}\\
		&=\frac{2x}{x^2\cdot g^{(2)}_{S}+g^{(2)}_{E}+2x},
	\end{aligned}
\end{equation}
 $x=P_S(1)/P_E(1)$ is the brightness ratio between the thermal field and auxiliary entangled field. Higher-order events from both fields add indistinguishable noise to the coincidence counts, degrading the measured visibility. We numerically evaluate this effect by plotting the ideal visibility against the field ratio (see equation above), with each field's $g^{(2)}$. The calculations demonstrate: (1)For our auxiliary entanglement source's actual $g^{(2)}(0)=0.13\pm0.08$, the peak visibility occurs at a brightness ratio of 0.22, and the maximum visibility $V_H=0.69$. Guided by this result, we configure the brightness of thermal field to $\braket{n}=0.058$ under $\eta_{ro}\approx 0.26$ for maximal visibility. The 1-km fiber (about 3~dB loss) in the 5 $\upmu$s delayed arm attenuates the thermal field, inducing visibility variations as summarized in Tab.~\ref{tab:V_H}. (2) With an ideal single photon source, $g^{(2)}(0)=0$, the visibility could approach unity. This implies that, all other conditions being equal, lower stellar light brightness would yield higher visibility; (3) Coherent light as auxiliary source limits visibility to 0.41 and optimal at $1:\sqrt{2}$ brightness ratio. Under conditions of faint stellar light, $\braket{n}\ll1$, the achievable count rates—comparable to those of intensity interferometers—would severely limit the feasibility of this scheme.
\begin{table}[h]\label{tab:V_H}
	\centering
	\caption{A summary of $V_{\rm H}$.}
\begin{tabular}{C{3cm}C{3cm}C{3cm}}
	\toprule
      experiment setup& equal length & 1~km delay \\
	  \midrule
	  $V_{H}$& 0.69& 0.579 \\
	  \bottomrule
\end{tabular}
\end{table}
\subsection{Degradation of coherence}
The field coherence (Eq.~\ref{eq:visibility} in the main text, numerator $d\cdot g$ terms) directly modulates the interference visibility. The thermal field's coherence time fundamentally limits detection: a detection window wider than the optical field's coherence time causes the photon number distribution evolves from Bose-Einstein statistics to Poissonian statistics. This evolution, which signifies a loss of coherence, thereby directly degrades the interference visibility. While increased readout power can compress wavepackets below 60~ns without efficiency loss, we adopted 60~ns and 20~ns detection window width. The correlation $g^{(2)}(0)=1+\lvert g^{(1)}(0)\rvert^2$ then quantifies coherence-dependent visibility reduction in Tab.~\ref{tab:g2}
\begin{table}[h]\label{tab:g2}
	\centering
	\caption{The calculated $g^{(2)}(0)$ values under different window width and the corresponding $V_{C}$.}
	\begin{tabular}{C{3cm}C{3cm}C{3cm}}
		\toprule
		 window width&20~ns &60~ns \\
		\midrule
		$g^{(2)}$ & $1.84$ & $1.45$ \\
		\midrule
		$V_{C}$ & $ 0.917 $ & $0.67$ \\
		\bottomrule
	\end{tabular}
\end{table}

\subsection{Phase instability}
As discussed above, three interferometers need to be stabilized. Each interferometer is actively locked during atomic loading periods and passively stabilized during the 3-ms interference intervals. Consequently, the phase instability of each interferometer contributed to visibility reduction:
\begin{equation}
	V_P=\int_{-\infty}^{+\infty}f(\delta \theta) \cos(\delta \theta) d\delta\theta,
\end{equation}
we assume that $\delta \theta$ follows a Gaussian probability distribution with a mean of 0 and a standard deviation (SD) of $\sigma=\sqrt{\sum_{i=1}^{i=3}\sigma_i^2}$. Then, we will get:
\begin{equation}
	V_{P}=e^{-\sigma^2/2}.
\end{equation}
The phase instability of each interferometer and the visibility reduction $V_{P}$ are given in Tab.~\ref{tab:phase_instability}
\begin{table*}[h]
	\centering
	\caption{The phase instability of each interferometer and  $V_{P}$}
	\begin{tabular}[t]{C{5cm}C{5cm} C{3cm} C{3cm}}
		\toprule
		$\sigma_\mathrm{THI}$ (1~km delay $\sigma_\mathrm{THI}$)&$\sigma_\mathrm{WOI}$ (1~km delay $\sigma_\mathrm{WOI}$)&$\sigma_\mathrm{WRI}$ &$V_{\rm P}$~(1~km delay $V_{\rm P}$)  \\
		\midrule
   $0.043$~rad~($0.209$~rad) & $0.063$~rad~(0.281~rad) & $0.081$~rad &$0.993$ ~($0.937$)\\
   \bottomrule
	\end{tabular}
	\label{tab:phase_instability}
\end{table*}
\subsection{Mismatch of two light field}
While the derived visibility formula assumes ideal field overlap, experimental implementations suffer from mode mismatch between independent sources. We characterize this through Hong-Ou-Mandel-type indistinguishability measurements between read-out fields and thermal fields~\cite{tsujimoto2021,liu2024}
\begin{equation}
	g_{\rm HOM}^{(2)}=\frac{g_\mathrm{th}^{(2)}+\zeta^2g_\mathrm{ro}^{(2)} +2(1-\eta)\zeta}{(1+\zeta)^2},
\end{equation}
where $\zeta$ is the relative intensity of two fields, and we set $\zeta=1$. The measured $g^{(2)}$, $\eta$ and the value $V_I$ that describes the decrease of V are listed at Tab.~\ref{tab:indistinguishability}.
\begin{table*}[ht]\label{tab:indistinguishability}
	\centering
	\caption{Results of measured $g$, $\eta$ and $V_{I}$.}
	\begin{tabular}[t]{C{3cm}C{3cm} C{3cm} C{3cm} C{3cm}}
		\toprule
		$g^{(2)}_{\rm HOM_{L}}$& $\eta_{\rm L}$&$g^{(2)}_{\rm HOM_{R}}$&$\eta_{\rm R}$&$V_{I}$ \\
		\midrule
		$0.997\pm0.035$&$0.93\pm0.07$ & $0.98\pm 0.04$& $0.96\pm0.08$&$0.94\pm0.05$\\
		\bottomrule
	\end{tabular}
\end{table*}

\subsection{Summary}
Finally, we summarize all experimental imperfections discussed above and visibility $V_I$ of assisted entanglement in Tab.~\ref{tab:result}. After multiplying all of them, we can get the theory value $V_{\rm theory}$ to compare with the experimental results. Visibility is primarily limited by higher-order excitations and thermal-field coherence degradation. All experiment results are a little lower than the theory value, which hints other uncaptured experimental imperfections.  

\begin{table*}[htb]\label{tab:result}
	\caption{The comparison between the theoretical estimation of $V$ and the experimental results.}
	\begin{tabular}[t]{C{3cm} C{3cm} C{3cm} C{3cm} C{3cm}}
		\toprule
		experiment setup&equal arm, local, 20~ns window&equal arm, 20~km, 20~ns window&equal arm, 20~km, 60~ns window&1~km delay, 20~km, 60~ns window\\
		\midrule
		$V_{\rm SNR}$&$0.998\pm0.030$&$0.994\pm0.030$&$0.994\pm0.030$ &$0.994\pm0.030$\\
		%\midrule
		$V_{H}$&$0.69$&$0.69$ & $0.69$&$0.576$\\
		%\midrule
		$V_{C}$&$0.917$& $0.917$&$0.67$ &$0.67$\\
		%\midrule
		$V_{P}$&$0.993$&$0.993$ & $0.993$&$0.937$\\
		%\midrule
		$V_{I}$&$0.94\pm0.05$&$0.94\pm0.05$ & $0.94\pm0.05$&$0.94\pm0.05$\\
		\midrule
		$V_{\rm theory}$&$0.59\pm0.04$&$0.59\pm0.04$&$0.429\pm0.025$&$0.338\pm0.021$\\
		%\midrule
		$V_{\rm exp}$&$0.54\pm0.05$&$0.51\pm0.04$&$0.39\pm0.01$ & $0.32\pm0.03$\\
		\bottomrule
	\end{tabular}
\end{table*}

\end{document}